\documentclass[submission,copyright,creativecommons]{eptcs}
\providecommand{\event}{HCVS 2026} %

\usepackage{iftex}

\usepackage{minted}

\usepackage{hyperref}
\usepackage{amsmath}
\usepackage{amssymb}
\usepackage{bussproofs}
\usepackage[dvipdfmx]{graphicx}
\usepackage{multirow}

\ifpdf
  \usepackage{underscore}         %
  \usepackage[T1]{fontenc}        %
\else
  \usepackage{breakurl}           %
\fi

\title{CHC-based Automated Verification of WebAssembly Programs}
\author{Akihisa Yagi \qquad\qquad Ken Sakayori \qquad\qquad Naoki Kobayashi
\institute{The University of Tokyo\\ Tokyo, Japan}
\email{\{akihisa-yagi, sakayori, koba\}@is.s.u-tokyo.ac.jp}}
\def\titlerunning{CHC-based Automated Verification of WebAssembly Programs}
\def\authorrunning{Akihisa Yagi, Ken Sakayori \& Naoki Kobayashi}
\begin{document}
\maketitle

\begin{abstract}
WebAssembly is a stack-based imperative language widely used to develop safe and efficient Web applications. In this paper, we propose an automated static verification method for a subset of WebAssembly using a constrained Horn clauses (CHCs) satisfiability solver. Our main challenges are how to handle indirect function calls effectively and how to analyze huge panic handlers.
A naïve approach to the former problem would be to model a function reference table as an array of functions' entry points, but it would suffer from having too many candidates for indirect calls, resulting in a large case analysis. We address the problem by utilizing type information and filtering candidates for each indirect function call.
For the latter problem, a panic handler, which is a function that is called when an error occurs, can be very large and complex. We mitigate this problem by summarizing the panic handler using control-flow analysis. We confirmed the effectiveness of our approach through preliminary experiments.
\end{abstract}

\section{Introduction}
\label{sec:intro}

WebAssembly~\cite{wasmcore} is a stack-based low-level language designed to provide a safe and efficient execution environment. The language plays a crucial role not only in the development of web applications but also in various execution environments. As it is increasingly used as a target for systems programming languages, verifying the safety of WebAssembly programs becomes an important task. In this paper, we propose a CHC-based verification method for such safety properties. We represent error states, such as assertion violations, by \texttt{unreachable} instructions, and verify safety by proving that such instructions are unreachable during execution.

Our method translates a given WebAssembly program into corresponding constrained Horn clauses (CHCs). The reachability verification problem is then reduced to the satisfiability problem of the generated CHCs. We can use existing CHC solvers such as Z3 Spacer~\cite{komuravelli2016smt} or Eldarica~\cite{hojjat2018eldarica} to automatically prove safety. In this respect, our approach is related to CHC-based verification of stack-based bytecode, such as JayHorn~\cite{kahsai2016jayhorn} for Java bytecode because WebAssembly is also a stack-based. Additionally, WebAssembly's validation phase guarantees that the shape and types of the operand stack are statically determined at each instruction. This allows us to translate stack operations without representing the operand stack explicitly as a list or an array in the generated CHCs.

At the same time, WebAssembly differs from Java bytecode in several
important respects, making a direct adaptation of existing approaches
insufficient. First, WebAssembly is low-level and makes extensive use of
bitwise operations. To reason precisely about such operations, the
generated CHCs should model values as bit vectors rather than as
unbounded integers. Second, WebAssembly supports indirect function calls
via function tables, which are used to implement various forms of dynamic
control flow in source languages and runtime systems. Therefore, indirect
function calls must be handled precisely rather than treated by an
imprecise over-approximation. Third, WebAssembly binaries tend to contain
a large amount of auxiliary code, such as runtime support code and panic
handlers, which is irrelevant to the safety properties of interest. A
na\"ive translation of the whole program can therefore produce
unnecessarily large CHCs and make verification ineffective.

In this paper, we focus on the latter two challenges: indirect function calls and auxiliary code that does not affect the safety properties being verified. Bit-vector operations can be represented directly using the corresponding theories supported by modern CHC solvers. Unlike bit-vector operations, these challenges are specific to the structure of WebAssembly programs and directly affect the size and shape of the generated CHCs.

We first address the treatment of \texttt{call\_indirect},
WebAssembly's instruction for indirect function calls. The instruction
calls a function reference stored at a specified index in a function
table, allowing the target function to be determined dynamically at
runtime. Although \texttt{call\_indirect} is conceptually an indirect
jump, WebAssembly restricts its behavior: the target must be a function
reference in a table, and its type must match the type expected at the
call site~\cite{wasmsecurity}. An invalid table access or a type mismatch
causes a trap, which we model as reaching an \texttt{unreachable}
instruction and thus treat as a safety violation.

These restrictions against \texttt{call\_indirect} makes indirect calls amenable to static verification. We use the expected function type at each call site to filter possible
callees, while still checking whether the call may fail due to an invalid
table access or a type mismatch. This avoids a case analysis over all
functions in the module and reduces the size of the generated CHCs. We
further introduce two optimizations based on properties of the function
reference table.

The other issue we address concerns panic handlers. In this paper, we use
the term \emph{panic handler} to refer to a function invoked when an
assertion fails or a runtime error occurs. Although panic handlers are not
part of WebAssembly itself, they commonly appear in WebAssembly programs
compiled from high-level languages. For example, WebAssembly programs
compiled from Rust often contain panic handlers consisting of a large
number of instructions. These instructions typically construct diagnostic
information and do not affect the safety property of interest. However,
if CHCs are generated for panic handlers directly, the resulting CHCs can
become too large and complex for existing solvers.

\begin{figure}
    \begin{center}
        \includegraphics[width=0.4\textwidth]{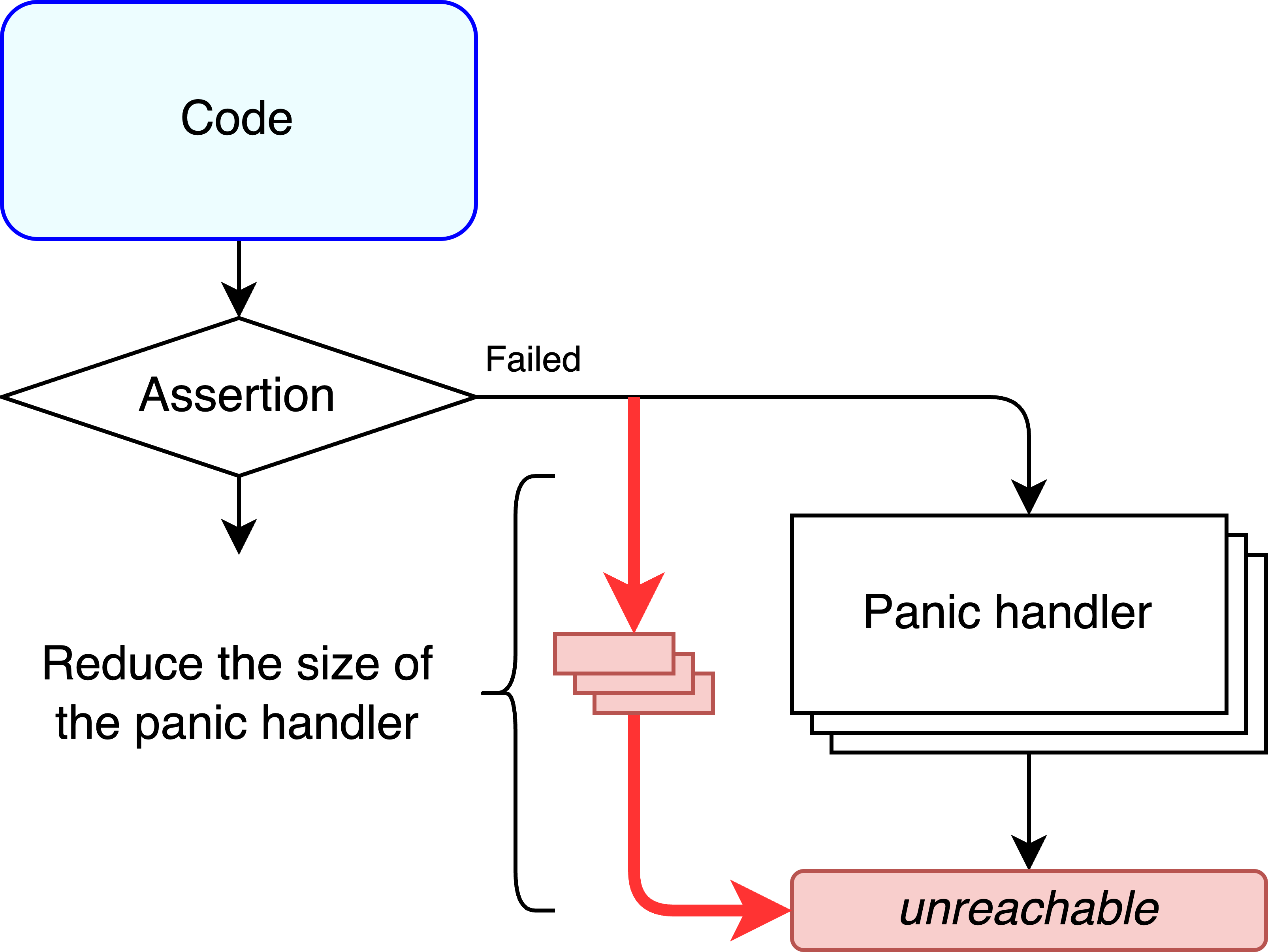}
    \end{center}
    \caption{An overview of our approach to handle panic handlers}
    \label{fig:panic_handler}
\end{figure}

To mitigate this problem, we slice panic handlers using control-flow
analysis, as shown in Figure~\ref{fig:panic_handler}. Since our
verification problem concerns reachability of \texttt{unreachable}
instructions, we can eliminate instructions that do not affect this
reachability. This technique does not require additional information
about panic handlers, such as their function names.

The rest of the paper is structured as follows.
Section~\ref{sec:target_language} defines the syntax and semantics of
the target language, an idealized subset of WebAssembly.
Section~\ref{sec:reduction} describes the reduction from the verification
problem for programs in the target language to CHC satisfiability, and
introduces two optimizations for handling \texttt{call\_indirect}
instructions.
Section~\ref{sec:panic_handlers} presents our technique for summarizing
panic handlers.
Section~\ref{sec:impl} introduces our verification tool
\textsc{WasmVerifier} and reports the results of preliminary experiments.
Section~\ref{sec:related_work} discusses related work, and
Section~\ref{sec:conclusion} concludes the paper.

\section{Target Language}
\label{sec:target_language}

We define the target language of our verification method, namely,
the language of programs to be verified. It is an idealized subset
of WebAssembly and is based on the formalization by Haas et al.~\cite{haas2017bringing}.

Like WebAssembly, our target language is a stack-based language: instructions implicitly operate on an operand stack. For example, an integer binary operation consumes two integer values from the stack and pushes the result back onto the stack. Thus, value operands are generally supplied through the operand stack, rather than being written explicitly in the instruction syntax. In addition to the operand stack, programs may use local variables, global variables, tables, and memories.

The syntax of the target language is shown below.

\newcommand{\tf}{t\!f}

\allowdisplaybreaks

\begin{align*}
    &\textrm{(value types)}\:t ::= \textrm{i32}\\
    &\textrm{(function type)}\:\tf ::= t^* \to t\;|\;t^* \to \varepsilon\\
    &\begin{aligned}
        \textrm{(instructions)}\:e ::= &\texttt{unreachable}\;|\;\texttt{nop}\;|\;\texttt{drop}\;|\;\texttt{select}\;|\\
        &\texttt{block}\;\tf\;e^*\;\texttt{end}\;|\;\texttt{loop}\;\tf\;e^*\;\texttt{end}\;|\;\texttt{if}\;\tf\;e^*\;\texttt{else}\;e^*\;\texttt{end}\;|\\
        &\texttt{br}\;i\;|\;\texttt{br\_if}\;i\;|\;\texttt{return}\;|\;\texttt{call}\;i\;|\;\texttt{call\_indirect}\;\tf\;i|\\
        &\texttt{get_local}\;i\;|\;\texttt{set\_local}\;i\;|\;\texttt{tee\_local}\;i\;|\;\texttt{get\_global}\;i\;|\texttt{set\_global}\;i\;|\\
        &\texttt{get\_table}\;i\;|\;\texttt{set\_table}\;i\;|\;t.\texttt{load}\;i\;|\;t.\texttt{store}\;i\;|\\
        &t.\texttt{const}\;c\;|\;t.unop|\;t.binop
    \end{aligned}\\
    &\begin{aligned}
        unop ::=&\:\texttt{eqz}\\
        binop ::=&\:\texttt{add}\:|\:\texttt{sub}\:|\:\texttt{mul}\:|\:\texttt{div\_s}\:|\:\texttt{div\_u}\:|\:\texttt{rem\_s}\:|\:\texttt{rem\_u}\\
        &|\:\texttt{and}\:|\:\texttt{or}\:|\:\texttt{xor}\:|\:\texttt{shl}\:|\:\texttt{shr_s}\:|\:\texttt{shr_u}\\
        &|\:\texttt{eq}\:|\:\texttt{ne}\:|\:\texttt{gt\_s}\:|\:\texttt{gt\_u}\:|\:\texttt{lt\_s}\:|\:\texttt{lt\_u}\:|\:\texttt{ge\_s}\:|\:\texttt{ge\_u}\:|\:\texttt{le\_s}\:|\:\texttt{le\_u}
    \end{aligned}\\
    &\textrm{(functions)}\:f ::= \texttt{func}\;\tf\;\texttt{local}\;t^*\;e^*\\
    &\textrm{(globals)}\:glob ::= \texttt{global}\;t\;e^*\\
    &\textrm{(tables)}\:tab ::= \texttt{table}\\
    &\textrm{(table elements)}\:el ::= \texttt{elem}\;n\;m\;i\\
    &\textrm{(memories)}\:mem ::= \texttt{memory}\\
    &\textrm{(program)}\:P ::= glob^*\;tab^*\;el^*\;mem^*\;f^*
\end{align*}

$t^*$ and \(e^*\) denote sequences of value types and instructions, respectively. We also use \(E\) to denote a sequence of instructions. We describe the syntax in more detail below.

\paragraph{Types.} For simplicity, we only consider 32-bit integer values. We omit floating-point values, packed values, vector values, global types, and functions with multiple return values. These features are orthogonal to the main issues addressed in this paper.

\paragraph{Instructions.} The language includes the main classes of WebAssembly instructions that are relevant to our CHC-based verification: structured control-flow instructions, branch instructions, direct and indirect function calls, local and global variable accesses, table operations, memory accesses, and integer operations.

Among these features, indirect calls and table operations are especially
important for our purpose. WebAssembly programs compiled from low-level
or systems languages often use function tables, and such tables may be
modified at runtime. We therefore include \texttt{get\_table} and
\texttt{set\_table}, in addition to \texttt{call\_indirect}. Since we
allow multiple function tables, the instruction \texttt{call\_indirect}
carries a table index \(i\).

We omit several WebAssembly instructions that are not essential for the
safety properties considered in this paper. For example, we omit
\texttt{br\_table}, which performs an indirect branch based on a table of branch targets, since it can be expanded into a sequence of
conditional branches. We also omit memory-size instructions such as
\texttt{current\_memory} and \texttt{grow\_memory}; instead, we assume
that memories are arbitrarily sized and do not model out-of-bounds
traps. If such traps need to be verified, they can be handled by adding
variables representing table and memory sizes and generating
corresponding bounds-checking constraints. We also omit exception-handling
instructions such as \texttt{throw} and \texttt{try\_table}.

\paragraph{Program.} A program consists of global declarations, table declarations, table
elements, memory declarations, and function declarations. The language
allows multiple tables and memories. The keywords \texttt{table} and
\texttt{memory} declare a table and a memory, respectively. Unlike actual
WebAssembly, we omit their initial sizes. For memories, we assume that
they are arbitrarily sized and initialized before the function to be
verified is called.

We assume that programs in the target language are valid in the sense of
WebAssembly validation. In particular, stack values have types consistent
with the instructions that consume them, and indices of functions, local
variables, globals, tables, memories, and branch targets are valid. This
assumption allows us to avoid modeling ill-typed executions and invalid
indices in the CHC translation.

Unless specified explicitly by declarations, table and memory entries are
assumed to be initialized nondeterministically. Thus, values loaded from
memory are not guaranteed to satisfy any particular property, and are
treated as part of the verification problem.

\section{Reduction to CHCs}
\label{sec:reduction}

As discussed in the introduction, we verify programs in the target
language by reducing their verification problem to constrained Horn
clause (CHC) solving, a standard framework for program verification
\cite{bjorner2015horn}. This section presents the reduction.

\subsection{Preparing Predicates}

Assume that each instruction in the target program is assigned a unique label. For each labeled instruction, we introduce a
predicate that summarizes the executions starting from that program
point. The predicate associated with an instruction labeled \(l\) in a
function \(f\) has the following form:
\begin{align*}
    f^l(\underbrace{S, H_L, H_G, H_T, H_M}_{\text{The current state}}, \underbrace{v_\text{ret}, H_G', H_T', H_M', v_\text{safe}}_{\text{The state after a function call}})
\end{align*}

The first group of arguments represents the program state at the
instruction. Here, \(S\) is the operand stack, \(H_L\) is the local store,
and \(H_G\), \(H_T\), and \(H_M\) represent globals, tables, and memories,
respectively. The remaining arguments summarize the result of executing
the current function from that point. The value \(v_{\text{ret}}\) is
the return value of the function, and \(H_G'\), \(H_T'\), and \(H_M'\)
are the globals, tables, and memories after the function returns. The
last argument \(v_{\text{safe}}\) is a Boolean value indicating whether
the execution is safe. If \(v_{\text{safe}} = 0\), the execution may
reach an \texttt{unreachable} instruction or a \texttt{call\_indirect} instruction that causes a type-mismatch trap. Since the operand stack
and local variables are no longer live after the function returns, the
post-state does not include \(S\) or \(H_L\).

Tables and memories are modeled as \texttt{Array}s in the generated
CHCs. Although it would also be natural to model the operand stack and
local variables as arrays, arrays are generally
difficult for CHC solvers to handle. We therefore use a flat encoding
for these components. WebAssembly validation determines the stack height
at each instruction, and the numbers of local and global variables are
fixed by the program. This allows these components to be represented by
separate \texttt{BV} arguments rather than by arrays.

Accordingly, if the stack height at label \(l\) is \(n\), the number of local
variables is \(m\), and the number of global variables is \(k\), then the
predicate \(f^l\) has the following type:
\begin{align*}
    f^l: &\overbrace{\texttt{BV}\times \cdots \times \texttt{BV}}^{\text{$n$ times}}\times \overbrace{\texttt{BV}\times \cdots \times \texttt{BV}}^{\text{$m$ times}} \times \overbrace{\texttt{BV}\times \cdots \times \texttt{BV}}^{\text{$k$ times}} \times \texttt{Array} \times \texttt{Array}\\
    &\times R \times \overbrace{\texttt{BV}\times \cdots \times \texttt{BV}}^{\text{$k$ times}} \times \texttt{Array} \times \texttt{Array} \times \texttt{BV} \to \texttt{Prop}
\end{align*}

Here, \(R\) is \texttt{BV} if \(f\) returns a value. If \(f\) has no
return value, the argument \(v_{\text{ret}}\) and the corresponding sort \(R\) are omitted.

\subsection{Generating CHCs}

We define rules for generating CHCs that encode the safety property of the program using these predicates. Each instruction gives rise to clauses describing the relation between the predicate associated with the instruction and the predicate associated with the next instruction. We implicitly quantify all free variables in the generated clauses universally.

Let $\mathcal{C}$ denote a function that converts a single instruction into a set of CHCs. To support instructions such as \texttt{br} and \texttt{br\_if}, which refer to branch labels outside the current instruction, the conversion function takes both an instruction and a stack of branch labels $B$ as arguments. The stack $B$ records the labels currently available for branch instructions, so that the target of the $i$th branch label is given by $B[i]$.

We now define $\mathcal{C}$. In the following rules, arguments of predicates that are not changed by the instruction are omitted. We write $(t :=) v$ to indicate that the argument $t$ is updated to $v$ by the instruction. The label $l$ denotes the label of the current instruction, and $\hat{l}$ denotes the label of the next instruction. For control instructions, the branch label stack $B$ is updated when translating their nested instructions. The variable $v_\text{safe}$ is represents whether the execution is safe: it has value $1$ for safe executions, and is set to $0$ when an \texttt{unreachable} instruction is reached or when a type-mismatch trap occurs in a \texttt{call\_indirect} instruction.

\begin{align*}
    &\begin{aligned}
        \mathcal{C}(\texttt{unreachable}, B) = &\{f^{\hat{l}}(\cdots, (v_\text{safe}:=)0)\leftarrow \top\}
    \end{aligned}\\
    &\begin{aligned}
        \mathcal{C}(\texttt{nop}, B) = &\{f^l(\cdots) \leftarrow f^{\hat{l}}(\cdots)\}
    \end{aligned}\\
    &\begin{aligned}
        \mathcal{C}(\texttt{drop}, B) = &\{f^l((S := )s :: S, \cdots) \leftarrow f^{\hat{l}}((S := )S, \cdots)\}
    \end{aligned}\\
    &\begin{aligned}
        \mathcal{C}(\texttt{select}, B) = &\{f^l((S := )s :: s_0 :: s_1 :: S, \cdots) \leftarrow f^{\hat{l}}((S := )s_0 :: S, \cdots) \land s = 0 \}\\
        \cup&\{f^l((S := )s :: s_0 :: s_1 :: S, \cdots) \leftarrow f^{\hat{l}}((S := )s_1 :: S, \cdots) \land s \neq 0 \}
    \end{aligned}\\
    &\begin{aligned}
        \mathcal{C}(\texttt{block}\;\tf\;E\:\texttt{end}, B) = &\{f^l(\cdots) \leftarrow f^{l_0}(\cdots)\} \cup \bigcup_{e \in E} \mathcal{C}(e, \hat{l} :: B)
    \end{aligned}\\
    &\begin{aligned}
        \mathcal{C}(\texttt{loop}\;\tf\;E\:\texttt{end}, B) = &\{f^l(\cdots) \leftarrow f^{l_0}(\cdots)\} \cup \bigcup_{e \in E} \mathcal{C}(e, l :: B)
    \end{aligned}\\
    &\begin{aligned}
        \mathcal{C}(\texttt{if}\;\tf\;E_0\:\texttt{else}\:&E_1\:\texttt{end}, B) =\\
        &\{f^l((S := ) s :: S, \cdots) \leftarrow f^{l_0}((S := ) S,\cdots) \land s \neq 0\}\\
        \cup&\{f^l((S := ) s :: S, \cdots) \leftarrow f^{l_1}((S := ) S,\cdots) \land s = 0\}\\
        \cup&\:\bigcup_{e \in E_0} \mathcal{C}(e, B)\cup \bigcup_{e \in E_1} \mathcal{C}(e, B)
    \end{aligned}\\
    &\begin{aligned}
        \mathcal{C}(\texttt{br}\:i, B) = &\{f^l(\cdots) \leftarrow f^{B[i]}(\cdots)\}
    \end{aligned}\\
    &\begin{aligned}
        \mathcal{C}(\texttt{br\_if}\:i, B) = &\{f^l((S := )s :: S, \cdots) \leftarrow f^{B[i]}((S := ) S, \cdots) \land s \neq 0\}\\
        \cup&\{f^l((S := )s :: S, \cdots) \leftarrow f^{\hat{l}}((S := ) S,\cdots) \land s = 0\}
    \end{aligned}
\end{align*}

Function calls require special care, because safety of a continuation
depends on whether the callee returns safely. Consider a sequence of
instructions \((\texttt{call}\:i)::E\). If the call to \(f_i\)
returns safely, then the continuation \(E\) is executed from the
state returned by \(f_i\). If the call to \(f_i\) is unsafe, then the
whole execution is unsafe, independently of the continuation. This is encoded by the following clauses:

\begin{align*}
    \mathcal{C}(\texttt{call}\:i, B) &=\left\{\begin{aligned}
        &f^l(\text{args}(f_i) @ S, H_L, H_G, H_T, H_M, v_\text{ret}, H_G', H_T', H_M', v_\text{safe})\\
        &\leftarrow f_i(\text{args}(f_i), H_G, H_T, H_M, v_\text{ret}', H_G'', H_T'', H_M'', 1)\\
        &\hspace{0.5cm}\land f^{\hat{l}}(v_\text{ret}' :: S, H_L, H_G'', H_T'', H_M'', v_\text{ret}, H_G', H_T', H_M', v_\text{safe})
    \end{aligned}\right\}\\
    &\cup \left\{\begin{aligned}
        &f^l(\text{args}(f_i) @ S, H_L, H_G, H_T, H_M, v_\text{ret}, H_G', H_T', H_M', 0)\\
        &\leftarrow f_i(\text{args}(f_i), H_G, H_T, H_M, v_\text{ret}', H_G'', H_T'', H_M'', 0)
    \end{aligned}\right\}
\end{align*}

The first clause handles the case where the call to \(f_i\) returns
safely. The predicate \(f_i\) is invoked with the arguments popped from
the operand stack, together with the current globals, tables, and
memories. Its last argument is fixed to \(1\), indicating that the
callee execution is safe. The continuation predicate \(f^{\hat{l}}\) is
then applied to the state returned by \(f_i\): the return value
\(v_\text{ret}'\) is pushed onto the stack, and the globals, tables, and
memories are updated to \(H_G''\), \(H_T''\), and \(H_M''\).

The second clause handles the case where the call to \(f_i\) is unsafe.
In that case, the whole execution from label \(l\) is unsafe, and the
continuation need not be considered. The variables
\(H_G''\), \(H_T''\), and \(H_M''\) denote the intermediate state after
the call to \(f_i\); like the other variables in the clauses, they are
implicitly universally quantified.

\begin{align*}
    &f^l(\underbrace{\text{args}(f_i) @ S, H_L, H_G, H_T, H_M}_{\text{The current state}}, \underbrace{v_\text{ret}, H_G', H_T', H_M', v_\text{safe}}_{\text{The state after $f$ is called}})\\
    &\leftarrow f_i(\underbrace{\text{args}(f_i), H_G, H_T, H_M}_{\text{The current state passed to $f_i$}}, \underbrace{v_\text{ret}', H_G'', H_T'', H_M'', 1}_{\text{The state after $f_i$ is called}})\\
    &\hspace{0.5cm}\land f^{\hat{l}}(\underbrace{v_\text{ret}' :: S, H_L, H_G'', H_T'', H_M''}_{\text{The state after $f_i$ is called}}, \underbrace{v_\text{ret}, H_G', H_T', H_M', v_\text{safe}}_{\text{The state after $f$ is called}})
\end{align*}

The instruction \texttt{call\_indirect} is handled similarly to
\texttt{call}, but the callee is determined by a function index obtained
from a table. We must also account for the case where the function
selected from the table does not have the expected function type, in
which case the instruction traps. Let \(\tf\) be the expected function
type, and let \(F(\tf)\) be the set of indices of functions whose type is
\(\tf\). Since the target language does not allow functions to be
created at runtime, the finite set \(F(\tf)\) can be computed
statically.

\begin{align*}
    \mathcal{C}&(\texttt{call\_indirect}\:\tf\;i, B) =\\
    &\bigcup_{m \in F(\tf)}\left\{\begin{aligned}
        &f^l(s :: \text{args}(f_m) @ S, H_L, H_G, H_T, H_M, v_\text{ret}, H_G', H_T', H_M', v_\text{safe})\\
        &\leftarrow f_m(\text{args}(f_m), H_G, H_T, H_M, v_\text{ret}', H_G'', H_T'', H_M'', 1)\\
        &\hspace{0.5cm}\land f^{\hat{l}}(v_\text{ret}' :: S, H_L, H_G'', H_T'', H_M'', v_\text{ret}, H_G', H_T', H_M', v_\text{safe})\\
        &\hspace{0.5cm}\land (H_T)_i[s] = m
    \end{aligned}\right\}\\
    \cup&\bigcup_{m \in F(\tf)}\left\{\begin{aligned}
        &f^l(s :: \text{args}(f_m) @ S, H_L, H_G, H_T, H_M, v_\text{ret}, H_G', H_T', H_M', 0)\\
        &\leftarrow f_m(\text{args}(f_m), H_G, H_T, H_M, v_\text{ret}', H_G'', H_T'', H_M'', 0)\\
        &\hspace{0.5cm}\land (H_T)_i[s] = m
    \end{aligned}\right\}\\
    \cup&\left\{\begin{aligned}
        &f^l(s :: S, H_L, H_G, H_T, H_M, v_\text{ret}, H_G', H_T', H_M', 0)\\
        &\leftarrow \bigwedge_{m \in F(\tf)} (H_T)_i[s] \neq m
    \end{aligned}\right\}
\end{align*}

The first two groups of clauses correspond to the case where the table
entry \((H_T)_i[s]\) is a function index \(m\) whose type matches the
expected type \(\tf\). As in the case of \texttt{call}, the first group
handles safe executions of the callee, while the second group propagates
unsafety of the callee to the current function. The additional constraint
\((H_T)_i[s]=m\) connects the function selected from the table with the
callee predicate \(f_m\).

The third clause covers the type-mismatch case: if the table entry is
not any function index in \(F(\tf)\), then the indirect call traps, and
the execution from label \(l\) is unsafe. Since \(F(\tf)\) is finite,
the conjunction \(\bigwedge_{m\in F(\tf)} (H_T)_i[s]\neq m\) is a finite
constraint, so the generated formula is still a CHC.

For a \texttt{return} instruction, the current globals, tables, and
memories become the post-state of the current function. Thus, the CHC
for \texttt{return} is generated as follows:

\begin{align*}
    &\begin{aligned}
        \mathcal{C}(\texttt{return}, B) = &\{f^l(s :: S, H_L, H_G, H_T, H_M, s, H_G, H_T, H_M, 1) \leftarrow \top\}
    \end{aligned}
\end{align*}

If the current function has no return value, the return-value argument
\(s\) and the corresponding argument \(v_{\text{ret}}\) are omitted.
If the function body does not end with an explicit \texttt{return}
instruction, we add an implicit \texttt{return} at the end of the body.
The CHCs for the remaining instructions are generated similarly; the
full set of rules is given in Appendix~\ref{sec:appendix-chc}.

We conjecture that this reduction is sound and complete. A formal proof of these properties has not yet been established, as this work is still in progress. We leave such a proof to future work.

\subsection{Optimizations on Indirect Function Calls}

We now describe two optimizations for the CHCs generated from
\texttt{call\_indirect}. The naive translation given above introduces
constraints over \texttt{Array}s for accessing function reference
tables. Such constraints can make the generated CHCs harder for solvers
to handle. The optimizations below reduce the use of array operations
under common restrictions on function reference tables.

The first optimization applies when the function reference table is
read-only. Even read-only tables are useful: they can be used to encode
case distinctions on function calls, or to represent integer values as
function pointers. Since a read-only table \((H_T)_i\) does not change during
execution, we can statically compute the indices at which it stores each function index, and expand \texttt{call\_indirect} instructions to case distinctions on the index \(s\) read from the table. For a function index \(m\), let \(I((H_T)_i, m)\)
denote the set of indices \(n\) such that \((H_T)_i[n]=m\). Then the
clauses for \texttt{call\_indirect} can be optimized as follows:

\begin{align*}
    \mathcal{C}&(\texttt{call\_indirect}\;\tf\;i, B) =\\
    &\bigcup_{m \in F(\tf)}\left(\bigcup_{n \in I((H_T)_i, m)}\left\{\begin{aligned}
        &f^l(s :: \text{args}(f_m) @ S, \cdots, v_\text{safe})\\
        &\leftarrow f_m(\text{args}(f_m), \cdots, 1)\land f^{\hat{l}}(v_\text{ret}' :: S, \cdots, v_\text{safe})\land s = n
    \end{aligned}\right\}\right)\\
    \cup&\bigcup_{m \in F(\tf)}\left(\bigcup_{n \in I((H_T)_i, m)}\left\{\begin{aligned}
        &f^l(s :: \text{args}(f_m) @ S, \cdots, 0)\\
        &\leftarrow f_m(\text{args}(f_m), \cdots, 0)\land s = n
    \end{aligned}\right\}\right)\\
    \cup&\left\{f^l(s :: S, \cdots, 0) \leftarrow \bigwedge_{m \in F(\tf)}\left(\bigwedge_{n \in I((H_T)_i, m)} s \neq n\right)\right\}
\end{align*}

Since the sets \(I((H_T)_i, m)\) are computed statically, the generated
clauses no longer contain array reads from the function reference table.

The second optimization applies when the table may be updated during
execution, but all accesses to it use constant indices. Such tables can
be regarded as a finite collection of mutable variables holding function
references, for example hooks. For each table entry \((H_T)_i[n]\) that
may be accessed, we introduce a fresh global variable \(h_{i,n}\),
represented as an additional component of \(H_G\). Let \(i_{i,n}\) be the
index of this component, so that \(H_G[i_{i,n}]\) represents the current
value of \((H_T)_i[n]\). Suppose that this occurrence of
\texttt{call\_indirect} accesses the \(n\)-th entry of table \(i\). Then
the clauses for \texttt{call\_indirect} are optimized as follows:

\begin{align*}
    \mathcal{C}&(\texttt{call\_indirect}\;\tf\;i, B) =\\
    &\bigcup_{m \in F(\tf)}\left\{\begin{aligned}
        &f^l(s :: \text{args}(f_m) @ S, \cdots, v_\text{safe})\\
        &\leftarrow f_m(\text{args}(f_m), \cdots, 1)\land f^{\hat{l}}(v_\text{ret}' :: S, \cdots, v_\text{safe})\land H_G[i_{i,s}] = m
    \end{aligned}\right\}\\
    \cup&\bigcup_{m \in F(\tf)}\left\{\begin{aligned}
        &f^l(s :: \text{args}(f_m) @ S, \cdots, 0)\\
        &\leftarrow f_m(\text{args}(f_m), \cdots, 0)\land H_G[i_{i,s}] = m
    \end{aligned}\right\}\\
    \cup&\left\{f^l(s :: S, \cdots, 0)\leftarrow \bigwedge_{m \in F(\tf)} H_G[i_{i,s}] \neq m\right\}
\end{align*}

This optimization affects the entire set of generated CHCs, because
every occurrence of a table read \((H_T)_i[n]\) must be replaced by the
corresponding global variable \(H_G[i_{i,n}]\). Unlike the optimization
for read-only tables, this transformation may increase the number of
global variables significantly, and hence may increase the size or
complexity of the generated CHCs.

Finally, if both conditions hold, that is, if the table is read-only and
all accesses to it use constant indices, then the callee of each
\texttt{call\_indirect} instruction can be determined statically. In
this case, we simply replace the instruction by the corresponding
\texttt{call} instruction.

\section{Optimization for Panic Handlers}
\label{sec:panic_handlers}

We call a function a panic handler if it is invoked when an
unrecoverable error occurs. In this section, we focus on panic handlers
that end with an \texttt{unreachable} instruction. This assumption is
reasonable for programs generated by compilers, because an unrecoverable
error corresponds to a program point that should not be reached during
normal execution. In practice, WebAssembly programs compiled from Rust
often contain large panic handlers that end with an \texttt{unreachable}
instruction.

A direct translation of such a panic handler may generate unnecessarily
large CHCs. This is because a panic handler often contains many
instructions for constructing diagnostic information, even though those
instructions are irrelevant to the safety property considered in this
paper: reachability of \texttt{unreachable}. Once execution enters such a
panic handler, the essential fact is that it eventually reaches
\texttt{unreachable}; the intermediate computations do not matter for the
verification of this property.

\begin{figure}[htb]
    \begin{center}
        \includegraphics[width=0.8\textwidth]{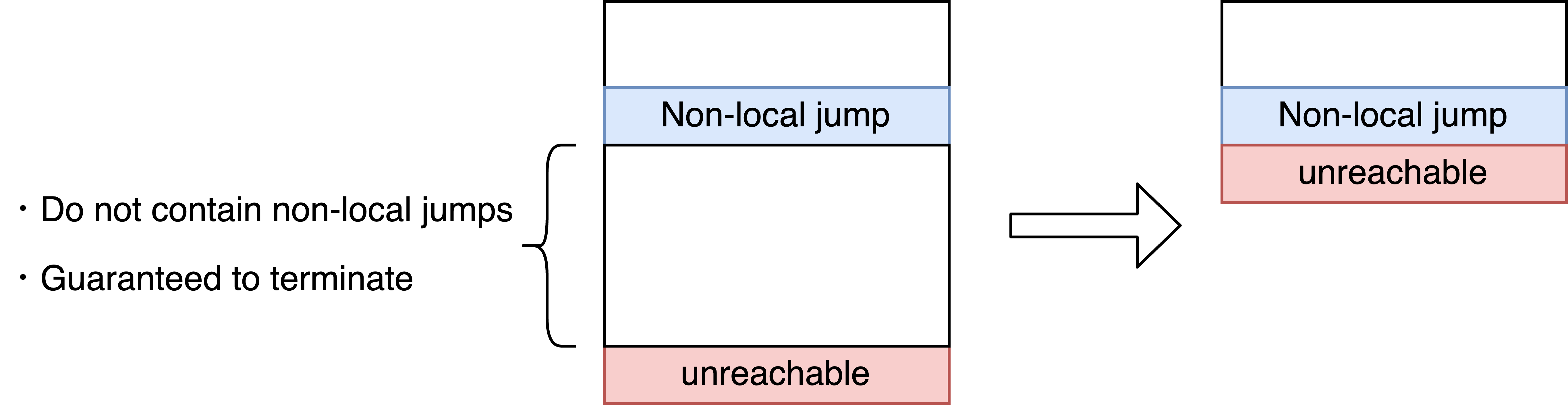}
    \end{center}
    \caption{Overview of the optimization for panic handlers}
    \label{fig:handler_optim}
\end{figure}

To reduce the size of the generated CHCs, we simplify instruction
sequences that immediately precede \texttt{unreachable}. In particular,
we repeatedly apply the following rewrite rule:

\begin{align*}
    e;\;\texttt{unreachable} &\rightarrow \texttt{unreachable}
\end{align*}

provided that the instruction \(e\) cannot transfer control to outside
the current continuation in a way that avoids the following
\texttt{unreachable}. Intuitively, this condition excludes instructions
such as \texttt{br}, \texttt{br\_if}, and \texttt{return}, which may skip
the subsequent \texttt{unreachable}. On the other hand, ordinary
computational instructions, local-variable instructions, and memory
instructions can be eliminated by this rule, since they do not affect
whether the following \texttt{unreachable} is reached.

The side condition of the rewrite rule can be checked by a simple
conservative control-flow analysis. For instance, if an instruction is
syntactically a \texttt{br}, \texttt{br\_if}, or \texttt{return}, we do
not eliminate it. For structured instructions such as \texttt{block},
\texttt{loop}, and \texttt{if}, we conservatively check whether their
bodies may contain such a control transfer. This conservative treatment
is sufficient for soundness, although a more precise control-flow
analysis could eliminate more instructions by determining that some
branches remain within the current scope.

Function calls require slightly more care. In our target language,
function calls do not perform non-local exits such as exceptions.
Therefore, from the viewpoint of control flow, a call does not skip the
following \texttt{unreachable}. However, the called function may diverge.
Eliminating such a call may turn a diverging execution into one that
reaches the following \texttt{unreachable}. This can only add spurious
reachability of \texttt{unreachable}; it never removes real reachability
present in the original program. Hence, the transformation is sound for
proving safety: if the optimized program is safe, then the original
program is also safe, although the optimization may introduce false
alarms.

\section{Implementation and Experiments}
\label{sec:impl}

We implemented the proposed method in a prototype tool called \textsc{WasmVerifier}.
The tool takes a WebAssembly binary or text file as input, applies the optimizations described in previous sections, translates it
into CHCs according to Section~\ref{sec:reduction}, and solves the generated CHCs using an external CHC solver. The tool
does not require any annotations on the input program. It automatically
and conservatively determines whether each optimization is applicable.

\textsc{WasmVerifier} currently supports only the instructions included
in the target language defined in Section~\ref{sec:target_language}. If the input
program contains an unsupported instruction, the tool replaces it with
\texttt{unreachable}. This treatment is conservative for safety
verification, but it may introduce false alarms in the verification
results.

\subsection{Experiments}

We conducted preliminary experiments to evaluate the effectiveness of our method.
The benchmark set consists of 13 original WebAssembly programs, 3 Rust programs
compiled to WebAssembly, and 74 C programs taken from the SV-COMP 2026
benchmarks~\cite{beyer_2026_sv_benchmarks}. For the Rust programs, we used
Rust 1.92.0 with the \texttt{wasm32-unknown-unknown} target to compile them
to WebAssembly. For the SV-COMP benchmarks, we selected C programs with the
\texttt{unreach-call} property from the \texttt{bitvector}, \texttt{loops},
and \texttt{recursive} categories, and compiled them to WebAssembly using
Clang 18.1.8. We excluded three C programs because \textsc{WasmVerifier}
could not load the generated WebAssembly files. The Wasm files in the \texttt{bitvector}, \texttt{loops}, and \texttt{recursive} benchmark sets had average sizes of 756, 549, and 502 bytes, respectively. The file sizes ranged from 213 to 1,511 bytes.

For each benchmark program, we measured the time taken to solve the CHCs
generated by \textsc{WasmVerifier}. Unsupported instructions were replaced
with \texttt{unreachable} instructions for the experiments. We ran the
experiments on a machine with an Intel Core i7-9700 processor and 48GB of RAM,
running Ubuntu 24.04. We used Z3 Spacer~\cite{komuravelli2016smt} 4.8.12 and
Eldarica~\cite{hojjat2018eldarica} 2.2.1 as CHC solvers. The timeout was set
to 300 seconds for each program.

\begin{figure}
    \begin{center}
        \begin{tabular}{|ll||r|r|r|r|}
        \hline
        Benchmarks & & \# & Passed & Timeout & Failed\\
        \hline
        \hline
        Original & & 13 & 10 & 3 & 0\\
        \hline
        Rust & & 3 & 1 & 1 & 1\\
        \hline
        \multirow{3}{*}{SV-benchmarks} & bitvector & 19 & 11 & 4 & 4\\
        & loops & 32 & 22 & 5 & 5\\
        & recursive & 23 & 12 & 9 & 2\\
        \hline
        \end{tabular}
    \end{center}
    \caption{Execution results on each benchmark with Z3 Spacer}
    \label{fig:exp_results}
\end{figure}

The results of the experiments are shown in Figure~\ref{fig:exp_results}.
With Z3 Spacer, \textsc{WasmVerifier} successfully verified or disproved
the safety properties of 56 out of 90 programs; it timed out on 22 programs
and reported false alarms on 12 programs. With Eldarica, it successfully
verified or disproved the safety properties of 54 programs; it timed out on
21 programs, reported false alarms on 11 programs, and failed due to
resource exhaustion in 4 cases: out of memory in 3 cases and stack overflow
in 1 case.

The false alarms were caused by our replacement of unsupported instructions
with \texttt{unreachable} instructions, which makes the generated CHCs
too imprecise to verify the safety properties of some target programs.

\section{Related Work}
\label{sec:related_work}

We discuss related work on verification of low-level programs, with a
particular focus on WebAssembly programs, and on CHC-based verification
of other programming languages.

\subsection{Low-level program verification}

wasm-verify~\cite{munuera2023specification} is a verification tool for
WebAssembly programs. It is based on symbolic execution and verifies
safety properties against specifications written in VerifiWasm, a
specification language designed for wasm-verify. The tool targets a
fragment of WebAssembly which does not include instructions for
operating on function reference tables or calling functions indirectly. In
contrast, our method does not require annotations or additional
specifications for the program to be verified, and supports indirect
function calls via function reference tables.

A different approach to WebAssembly verification is taken by
Iris-Wasm~\cite{rao2023iris}. Iris-Wasm is a higher-order separation
logic for WebAssembly, mechanized in Rocq, and is built on
Iris~\cite{jung2015iris}, a higher-order separation logic framework.
Iris-Wasm can be used to verify a wide range of properties of
WebAssembly programs and supports the \texttt{call\_indirect}
instruction. However, it provides a formalized program logic rather than
an automated verifier: users need to construct proofs for individual
programs in the logic. In contrast, our method is fully automated and
does not require manual proofs. Watt~\cite{watt2018mechanising} also
mechanizes the WebAssembly specification in Isabelle.

\subsection{CHC-based verification}

CHC solving has been established as a general framework for automated program
verification~\cite{bjorner2015horn}. Existing CHC-based techniques have been
used to verify imperative programs and programs written in or compiled to
languages such as Java, OCaml, C/LLVM IR, and Rust~\cite{toman2020consort,kahsai2016jayhorn,champion2020ice,gurfinkel2015seahorn,matsushita2021rusthorn}. We have already discussed JayHorn~\cite{kahsai2016jayhorn} in Section~\ref{sec:intro}. We discuss other representative work on CHC-based verification of low-level languages.

SeaHorn~\cite{gurfinkel2015seahorn} verifies programs represented in LLVM
IR~\cite{lattner2004llvm}. It is therefore applicable to languages that
compile to LLVM IR, especially C, and generates CHCs after preprocessing and
optimizing the LLVM IR. Although WebAssembly is also a low-level compilation
target, the target languages, however, have different execution models: LLVM IR is a
register-based instruction set architecture, whereas WebAssembly is a
stack-based language with statically validated stack behavior.

Bembenek et al.~\cite{bembenek2026bit} verify AArch64
programs by lifting them to Boogie, a verification-oriented intermediate
language, and then applying CHC-based verification. Their work is related to
ours in that both approaches use bit-vector theories to faithfully model
low-level operations. However, our method generates CHCs directly from
WebAssembly programs, rather than translating them through a separate
verification language such as Boogie. Moreover, the verification challenges are
different: AArch64 is a register-based instruction set architecture, whereas
WebAssembly is a stack-based language.

\section{Conclusion}
\label{sec:conclusion}

We proposed a CHC-based method for verifying reachability-based
safety properties of WebAssembly programs. The method translates
WebAssembly programs into CHCs while taking into account characteristic
features of WebAssembly programs, such as indirect function calls through
function reference tables.
To make the generated CHCs more amenable to existing CHC solvers, we
introduced optimizations for indirect calls that reduce the use of
\texttt{Array}-related operations, as well as a summarization technique
for panic handlers.

We implemented the proposed method in a prototype tool called
\textsc{WasmVerifier} and evaluated it on a set of benchmark programs.
The experimental results show that the tool can verify or disprove
reachability-based safety properties for a number of WebAssembly
programs. At the same time, the results also reveal several remaining
limitations: some benchmarks could not be handled because of timeouts,
memory exhaustion, false alarms caused by unsupported instructions, or
the difficulty of solving CHCs involving bit-vector operations.

These results indicate that CHC-based verification is a promising
approach to fully automated verification of WebAssembly programs, while
also suggesting directions for future work. In particular, further
improvements in CHC solving for bit-vector constraints, more precise
handling of currently unsupported WebAssembly instructions, and additional
optimizations in the translation to CHCs would be important for improving
the applicability of the method. In addition, although we
expect the reduction from WebAssembly programs to CHCs to be sound and
complete, a formal proof of these properties remains to be established.

\section*{Acknowledgements}

This work was supported by JST K Program Grant Number JPMJKP24U4, Japan.

\nocite{*}
\bibliographystyle{eptcs}
\bibliography{references}

\appendix

\section{CHC Generation Rule for Other Instructions}

\label{sec:appendix-chc}

\begin{align*}
    &\begin{aligned}
        \mathcal{C}(\texttt{get\_local}\:i, B) = &\{f^l(\cdots,(H_L)_i,\cdots) \leftarrow f^{\hat{l}}((S := ) (H_L)_i :: S, \cdots)\}
    \end{aligned}\\
    &\begin{aligned}
        \mathcal{C}(\texttt{set\_local}\:i, B) =&\{f^l((S := )s:: S,\cdots) \leftarrow f^{\hat{l}}(\cdots,((H_L)_i := ) s, \cdots)\}
    \end{aligned}\\
    &\begin{aligned}
        \mathcal{C}(\texttt{tee\_local}\:i, B) = &\{f^l((S := )s:: S,\cdots) \leftarrow f^{\hat{l}}((S := )s:: S,\cdots,((H_L)_i := ) s, \cdots)\}
    \end{aligned}\\
    &\begin{aligned}
        \mathcal{C}(\texttt{get\_global}\:i, B) = &\{f^l(\cdots,(H_G)_i,\cdots) \leftarrow f^{\hat{l}}((S := ) (H_G)_i :: S, \cdots)\}
    \end{aligned}\\
    &\begin{aligned}
        \mathcal{C}(\texttt{set\_global}\:i, B) =&\{f^l((S := )s:: S,\cdots) \leftarrow f^{\hat{l}}(\cdots,((H_G)_i := ) s, \cdots)\}
    \end{aligned}\\
    &\begin{aligned}
        \mathcal{C}(\texttt{get\_table}\:i, B)=
        &\left\{\begin{aligned}
            &f^l((S := )s :: S,\cdots,(H_T)_i,\cdots)\\
            &\leftarrow f^{\hat{l}}((S := ) (H_T)_i[s] :: S, \cdots)
        \end{aligned}\right\}
    \end{aligned}\\
    &\begin{aligned}
        \mathcal{C}(\texttt{set\_table}\:i, B) =
        &\left\{\begin{aligned}
            &f^l((S := )s_0 :: s_1 :: S,\cdots,(H_T)_i,\cdots)\\
            &\leftarrow f^{\hat{l}}(\cdots,((H_T)_i := )(H_T)_i[s_1 \mapsto s_0], \cdots)
        \end{aligned}\right\}
    \end{aligned}\\
    &\begin{aligned}
        \mathcal{C}(t.\texttt{load}\:i, B)=
        &\left\{\begin{aligned}
            &f^l((S := )s :: S,\cdots,(H_M)_i,\cdots)\\
            &\leftarrow f^{\hat{l}}((S := ) (H_M)_i[s] :: S, \cdots)
        \end{aligned}\right\}
    \end{aligned}\\
    &\begin{aligned}
        \mathcal{C}(t.\texttt{store}\:i, B) =
        &\left\{\begin{aligned}
            &f^l((S := )s_0 :: s_1 :: S,\cdots,(H_M)_i,\cdots)\\
            &\leftarrow f^{\hat{l}}(\cdots,((H_M)_i := )(H_M)_i[s_1 \mapsto s_0], \cdots)
        \end{aligned}\right\}
    \end{aligned}\\
    &\begin{aligned}
        \mathcal{C}(t.{const}\:c, B) = &\{f^l(\cdots) \leftarrow f^{\hat{l}}((S := )c :: S,\cdots)\}
    \end{aligned}\\
    &\begin{aligned}
        \mathcal{C}(t.unop, B) = &\{f^l((S := )s :: S\cdots) \leftarrow f^{\hat{l}}((S := )unop(s) :: S,\cdots)\}
    \end{aligned}\\
    &\begin{aligned}
        \mathcal{C}(t.binop, B) =&\{f^l((S := )s_0 :: s_1 :: S\cdots) \leftarrow f^{\hat{l}}((S := )binop(s_0, s_1) :: S,\cdots)\}
    \end{aligned}
\end{align*}

\end{document}